\documentclass[10pt,twocolumn,english,preprint,prb,tightenlines]{revtex4}
\usepackage[T1]{fontenc}
\usepackage[latin9]{inputenc}
\usepackage{amsmath}
\usepackage{graphicx}
\usepackage{amssymb}

\makeatletter

\providecommand*{\perispomeni}{\char126}
\AtBeginDocument{\DeclareRobustCommand{\greektext}{%
  \fontencoding{LGR}\selectfont\def\encodingdefault{LGR}%
  \renewcommand{\~}{\perispomeni}%
}}
\DeclareRobustCommand{\textgreek}[1]{\leavevmode{\greektext #1}}
\DeclareFontEncoding{LGR}{}{}

\providecommand{\tabularnewline}{\\}

\@ifundefined{textcolor}{}
{%
 \definecolor{BLACK}{gray}{0}
 \definecolor{WHITE}{gray}{1}
 \definecolor{RED}{rgb}{1,0,0}
 \definecolor{GREEN}{rgb}{0,1,0}
 \definecolor{BLUE}{rgb}{0,0,1}
 \definecolor{CYAN}{cmyk}{1,0,0,0}
 \definecolor{MAGENTA}{cmyk}{0,1,0,0}
 \definecolor{YELLOW}{cmyk}{0,0,1,0}
 }

\usepackage{amsfonts}
\providecommand{\U}[1]{\protect\rule{.1in}{.1in}}

\makeatother

\usepackage{babel}

\begin{document}

\title{\textmd{The Stabilization of Superconductivity by Magnetic Field
in Out-of-Equilibrium Nanowires}}

\author{Yu Chen, Yen-Hsiang Lin, S. D. Snyder, and A. M. Goldman}

\affiliation{School of Physics and Astronomy, University of Minnesota, Minneapolis,
MN 55455, USA}

\pacs{PACS number}
\begin{abstract}
A systematic study has been carried out on the previously reported
{}``magnetic-field-induced superconductivity'' of Zn nanowires.
By varying parameters such as magnetic field orientation and wire
length, the results provide evidence that the phenomenon is a nonequilibrium
effect associated with the boundary electrodes. They also suggest
there are two length scales involved, the superconducting coherence
length and quasiparticle relaxation length. As wire lengths approach
either of these length scales, the effect weakens. We demonstrate
that it is appropriate to consider the effect to be a stabilization
of superconductivity, that has been suppressed by an applied current. 
\end{abstract}
\maketitle
\volumeyear{2009} \volumenumber{number} \issuenumber{number}
\eid{identifier}

\date{{[}Date text{]}{date} \received{[}Received text{]}{date}\revised{[}Revised
text{]}{date}\accepted{[}Accepted text{]}{date}\published{[}Published
text{]}{date}}

\section{\textbf{\textup{\small Introduction}}}

Superconducting wires have potential for utilization in integrated
circuits, as a consequence of their dissipationless nature. Upon scaling
their sizes down below the coherence length, this characteristic can
be lost due to the destruction of superconducting long-range order
by either thermal or quantum fluctuations. Superconductors in this
quasi-one dimensional limit have nonzero resistances produced by phase
slip processes. This has been the focus of much research on superconducting
nanowires.\cite{Little1967,Langer1967,McCumber1970,Giordano1988,Zaikin1997,Newbower1972,Lau2001,Zgirski2005,Altomare2006}
The superconductivity of nanowires may be significantly influenced
by the state of their boundary electrodes. Because of the proximity
effect, one would expect an enhancement of superconductivity when
a wire is connected to superconducting electrodes, and a suppression
when connected to normal electrodes. These manifest themselves as
enhanced critical currents in superconducting microbridges\cite{Octavio1978}
and suppressed critical temperatures in Al nanowires with Cu-coated
Al electrodes.\cite{Boogaard2004} In addition, theoretical studies
have shown that a finite-length wire can undergo a superconductor-insulator
transition through its coupling to the external environment.\cite{Buchler2004,Fu2006}
Experimentally, a recent study of electro-deposited Zn nanowires found
that their coupling to bulk superconductors of other materials can
cause the so-called {}``anti-proximity effect''.\cite{Tian2005,Tian2006}
In contrast with the usual proximity effect, at certain temperatures
wires were found to enter the superconducting state from the normal
state when the electrodes were driven normal by a magnetic field.

A similar magneto-response was reported in a recent letter.\cite{Chen2009}
Lithographically-made Zn nanowires with Zn electrodes were found to
reenter the superconducting state upon the application of small magnetic
fields after being driven resistive by current at low temperatures.
Here, we report a systematic study of this effect, involving the variation
of several parameters. The results provide solid evidence that the
phenomenon is a nonequilibrium effect associated with the coupling
to the boundary electrodes. In addition, it is more appropriate to
treat it as a stabilization, or recovery, of superconductivity, which
was suppressed by the applied current.

The paper is structured in the following manner: previously reported
work on field enhanced superconductivity is reviewed in Section II.
Experiments on the field-orientation dependence of the effect are
presented in Section III. The wire length dependence is presented
in Section IV. Arguments that attribute the effect to the stabilization
of superconductivity are contained in Section V. Sections VI and VII
contain discussions of theories and conclusions, respectively.

\section{\textbf{\textup{\small Field-enhanced superconductivity}}}

As described in detail in Ref. 16, samples in the configuration of
a single Zn nanowire with wide Zn electrodes were prepared using a
combination of multi-layered photolithgraphy, electron-beam lithography
and vapor deposition. The last step of the fabrication process involves
evaporating Zn at a rate $\sim5\textrm{\AA/sec}$, with both the sample
and crucible shroud cooled to $77K$. This approach to wire fabrication
has the advantage of forming wires and electrodes simultaneously in
the same writing and deposition steps. It ensures a transparent interface
between the wire and the electrodes thereby maximizing the transport
of quasiparticles and Cooper pairs across the interface. After liftoff,
the samples were immediately transferred to a high-vacuum $^{3}$He
refrigerator where measurements were carried out using conventional
DC four-probe methods. Even though $I-V$ characteristics of wires
are non-linear except in their normal state, the resistance is calculated
as $R=V/I$, by averaging 25 measurements. This facilitates identification
of the three states of the wire that will be discussed subsequently.
\begin{table*}
\begin{centering}
\begin{tabular}{|c|c|c|c|c|c|c|c|}
\hline 
Sample & Width ($nm$) & Height ($nm$) & Length ($\mu m$) & $T_{c}$ (K) & $\xi(0K)$ $(\mu m)$ & $\rho_{Zn}(4.2K)$ (\textgreek{mW}-cm) & $I_{c}(0K)$ Theo. (\textgreek{m}A)\tabularnewline
\hline
\hline 
A & 85 & 150 & 1.5 & 0.85 & 0.17 & 11 & 120\tabularnewline
\hline 
B & 80 & 90 & 1.5 & 0.83 & 0.15 & 14 & 56\tabularnewline
\hline 
C & 60 & 100 & 1 & 0.76 & 0.26 & 8.4 & 53\tabularnewline
\hline 
D & 60 & 100 & 2 & 0.76 & 0.28 & 7.8 & 55\tabularnewline
\hline 
E & 60 & 100 & 4 & 0.76 & 0.31 & 6.3 & 61\tabularnewline
\hline 
F & 60 & 75 & 10 & 0.76 & 0.21 & 11 & 35\tabularnewline
\hline 
G & 65 & 100 & 1.5 & 0.78 & 0.31 & 7.6 & 63\tabularnewline
\hline
\end{tabular}
\par\end{centering}

\caption{Key parameters for several representative samples \label{Flo:table_param}}

\end{table*}

The effect is robust and has been observed consistently in more than
twenty samples. In Table \ref{Flo:table_param}, we list some key
parameters for several representative samples, which will be discussed
here. The wire widths and heights were determined by microscopy measurements.
The transition temperature $T_{c}$ was taken as the temperature of
the half-normal resistance at a low applied current of 0.1$\mu A$.
The zero-temperature dirty limit coherence length was estimated as
$\xi(0)\sim0.855\cdot\left(\xi_{0}l_{e}\right)^{1/2}$, where $\xi_{0}$
is the BCS coherence length, and $l_{e}$ is the mean free path that
is obtained from the product $\rho_{_{Zn}}l_{e}=2.2\times10^{-11}\Omega\cdot cm^{2}$
at 4.2 K. The value of $\rho_{_{Zn}}l_{e}$ used is from studies on
single crystal Zn nanowires.\cite{Schulz} For the wires of finite
lengths used in the current study, the conventional way of extracting
the coherence length from $H_{c}(T)$ near $T_{c}$ cannot be used,
due to the complications associated with the alteration of the boundary
conditions by the magnetic field. %
\begin{figure}
\begin{centering}
\includegraphics{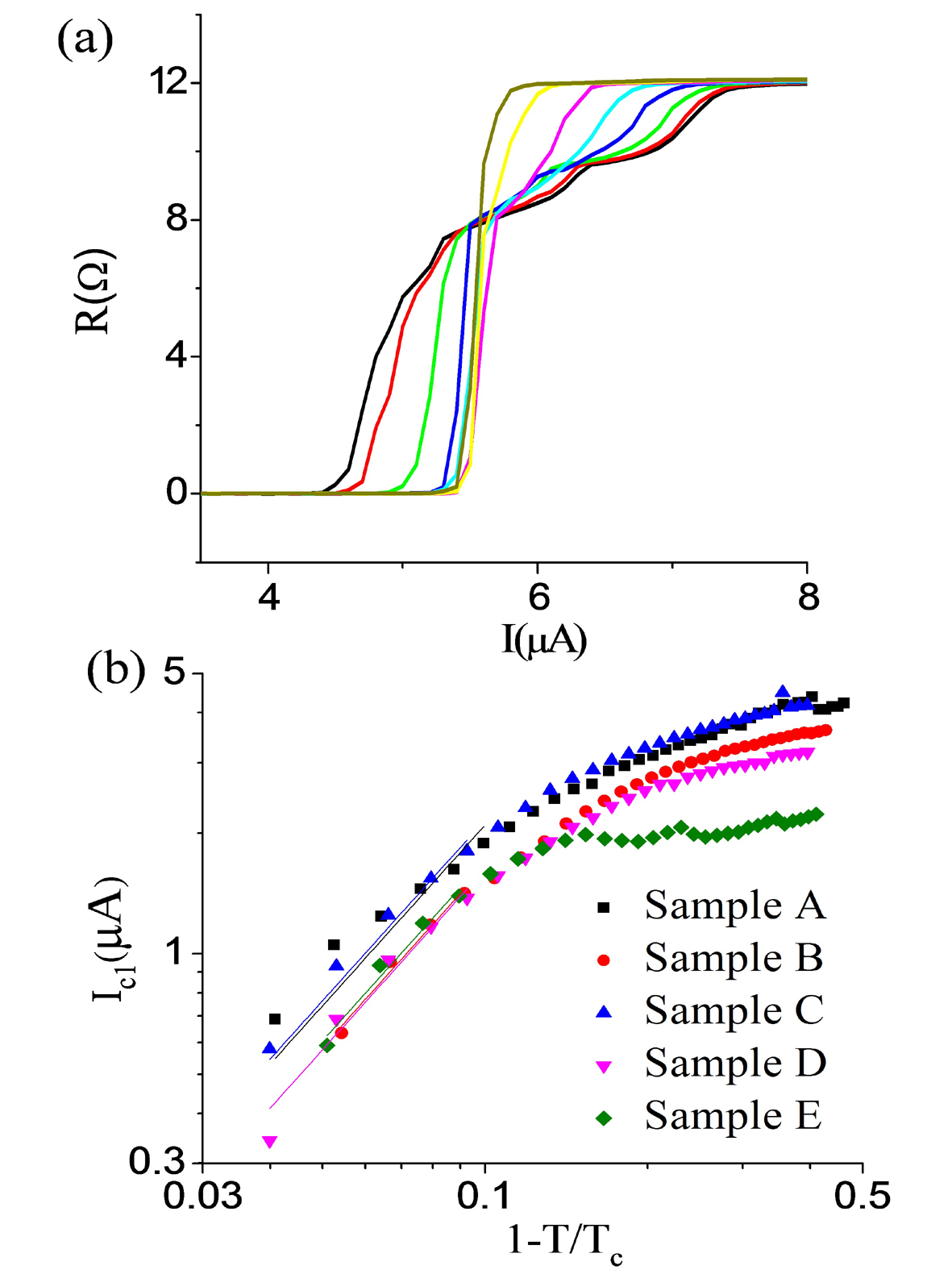}
\par\end{centering}

\caption{\textbf{\label{fig:RBI 1.5um}}(Color online) a) Current dependence
of the wire resistance (Sample A), at 460 mK, with the magnetic field
varying from 0 Oe to 28 Oe, at 4 Oe intervals. b) Log-log plot of
the critical current ($I_{c1}$) vs. temperature which is fit by the
GL theory (\ref{eq:1-1}) when $T/T_{c}>0.9$. The adjusted coefficient
of determination for each sample is: A = 0.827, B = 0.986, C = 0.967,
D = 0.968, and E = 0.973.}

\end{figure}

At low temperatures, the current-driven transition of the wire is
broad, associated with several characteristic currents. In the case
of a $1.5\mu m$ long sample (Sample A in Table I) shown in Fig. \ref{fig:RBI 1.5um}(a),
at $T=460mK$, the transition starts at $I_{c1}\backsimeq4.5\mu A$
with the onset of the non-zero resistance, stops at $I_{c2}\backsimeq7.7\mu A$
with the return to the normal resistance. In the transition regime,
there exists a shoulder-like structure developed around $I_{c0}\backsimeq5.5\mu A$.
(The other shoulder structure around $6.2\mu A$ is not a universal
feature and therefore will not be discussed here.) $I_{c1}$ has been
defined by the current at which $R/R_{n}>.01$ and is graphed as a
function of temperature in Fig.\ref{fig:RBI 1.5um}(b). As one can
see, there is agreement between the behavior of $I_{c1}$ and the
prediction of the Ginsburg-Landau (GL) theory for points with $T/T_{c}>0.9$,
but not over the entire temperature range.\cite{TinkhamBook} 

\begin{equation}
j_{c}=j_{c}(0)\left(1-\frac{T}{T_{c}}\right)^{3/2}\label{eq:1-1}\end{equation}
However, there is not agreement between this data and the Bardeen
expression

\begin{equation}
j_{c}=j_{c}(0)\left(1-\left(\frac{T}{T_{c}}\right)^{2}\right)^{3/2}\label{eq:1-2}\end{equation}
or Kupriyanov-Lukichev theory over the entire temperature window available.\cite{Bardeen1962,Romijn 1982}
Note that $I_{c1}$ is suppressed relative to either of these predictions
and that the transition in the $R$ vs $I$ curves is quite sharpe
near $T_{c}$ which makes $I_{c1}$, $I_{c0},$ and $I_{c2}$ almost
indistinguishable. An attempt to extract the temperature dependence
of $I_{c0}$ has been much more difficult as it is harder to define,
especially near $T_{c}$. In addition, $I_{c1}$, $I_{c0},$ and $I_{c2}$
deviate strongly from the theoretical predicted GL critical pair-breaking
current for isolated superconducting wires (see Table \ref{Flo:table_param}).\cite{Romijn 1982}
\begin{equation}
j_{c}(0)=\frac{8\pi^{2}\sqrt{2\pi}}{21\zeta(3)e}\left[\frac{(k_{B}T_{c})^{3}}{\hbar v_{f}\rho_{Zn}(\rho_{Zn}l_{e})}\right]^{1/2}\label{eq:1}\end{equation}
A part of the deviation may be associated with utilizing parameters
such as Fermi velocity of free electron model. The wires are also
appear granular under SEM and AFM scans which could lead to a system
of high disorder. This could explain the low critical currents.\cite{TinkhamBook}Another
explanation could be the oxidation layer at the surface of the nanowires
which could make the crossectional area smaller than our AFM and SEM
measurements or perhaps quantum confinement could explain it, but
it is all just speculation at this point. %
\begin{figure}
\includegraphics{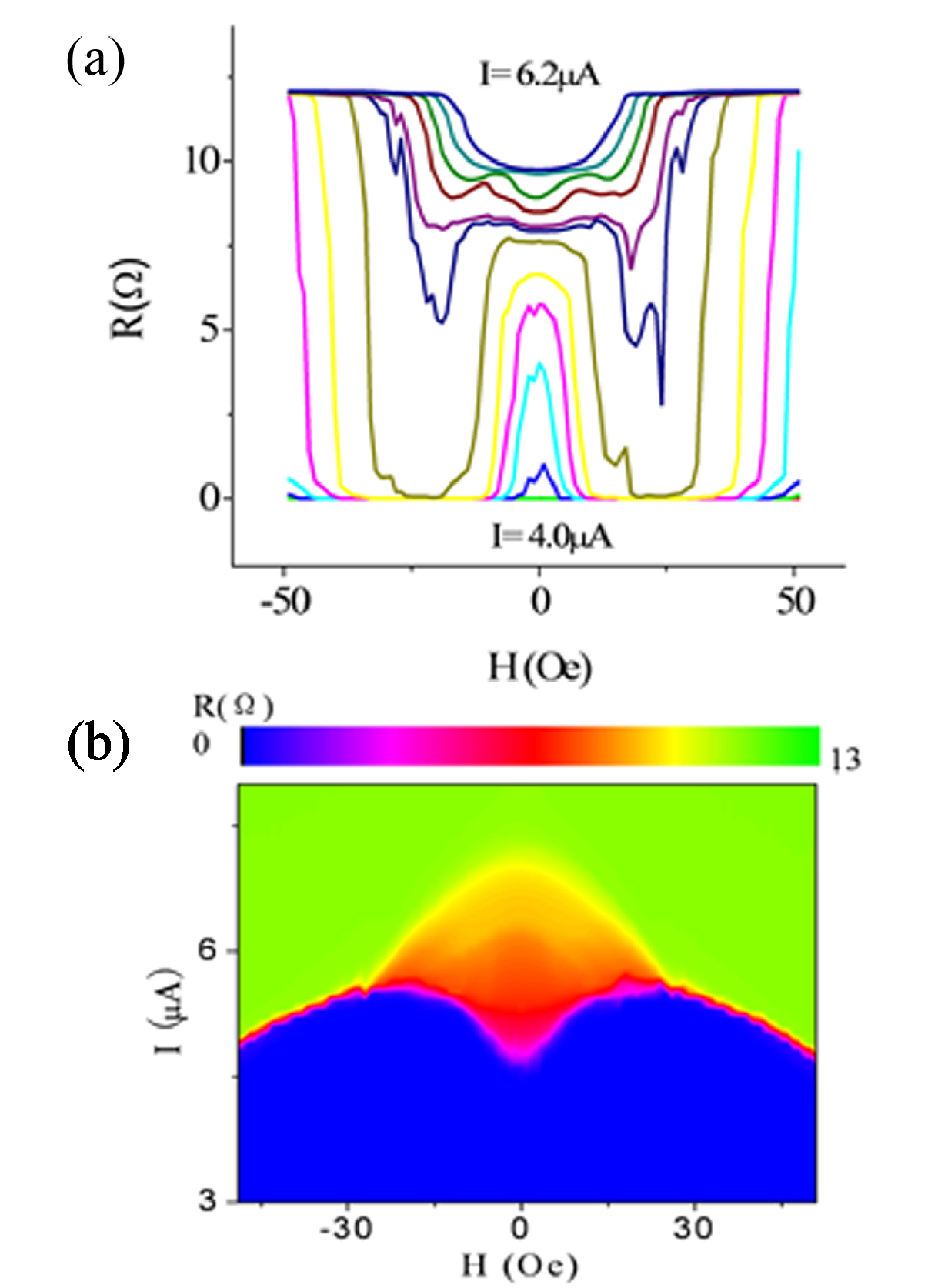}

\caption{\label{fig:Ic1(T)}(Color online) a) Magnetoresistance of this wire,
at currents from 4.0 $\mu A$ to 6.2 $\mu A$, at 0.2 $\mu A$ intervals.
b) False color phase diagram for this wire, at 460 mK.}

\end{figure}

When a magnetic field is applied, different parts of the transition
regime respond differently. Above the shoulder, the critical current
$I_{c2}$ decreases with increasing magnetic field. Below the shoulder,
the critical current $I_{c1}$ increases with increasing magnetic
field, until reaching a maximum of value equal $I_{c0}$. As a consequence,
over a range of currents, as shown in Fig. \ref{fig:Ic1(T)}(a), the
wire can re-enter its zero-resistance state from a current-driven
resistive state, upon the application of a small magnetic field. As
discussed in Ref. 16, a false color phase diagram of the wire can
be produced by graphing the wire resistance as a function of current
and magnetic field, as shown in Fig. \ref{fig:Ic1(T)}(b). The blue
region represents the zero-resistance state, the green region represents
the normal state and the colors other than these two represent the
intermediate resistive states of the transition region. In this phase
diagram, the enhancement of superconductivity is exhibited as the
V-shaped structure in the zero-resistance region, suggesting an increase
of the critical current $I_{c1}$ upon the application of a magnetic
field. These false color phase diagrams can also be created with a
temperature axis as opposed to a current axis. This was done in Ref.
16 and for two samples in this paper. Most of these plots are omitted
because it is rather redundant with the current data. One more experimental
observation worthy of mention is that the standard deviation of measured
resistances is much higher for those below the shoulder compared with
those above. Its origin remains unknown and is currently under investigation.
\begin{figure}
\begin{centering}
\includegraphics{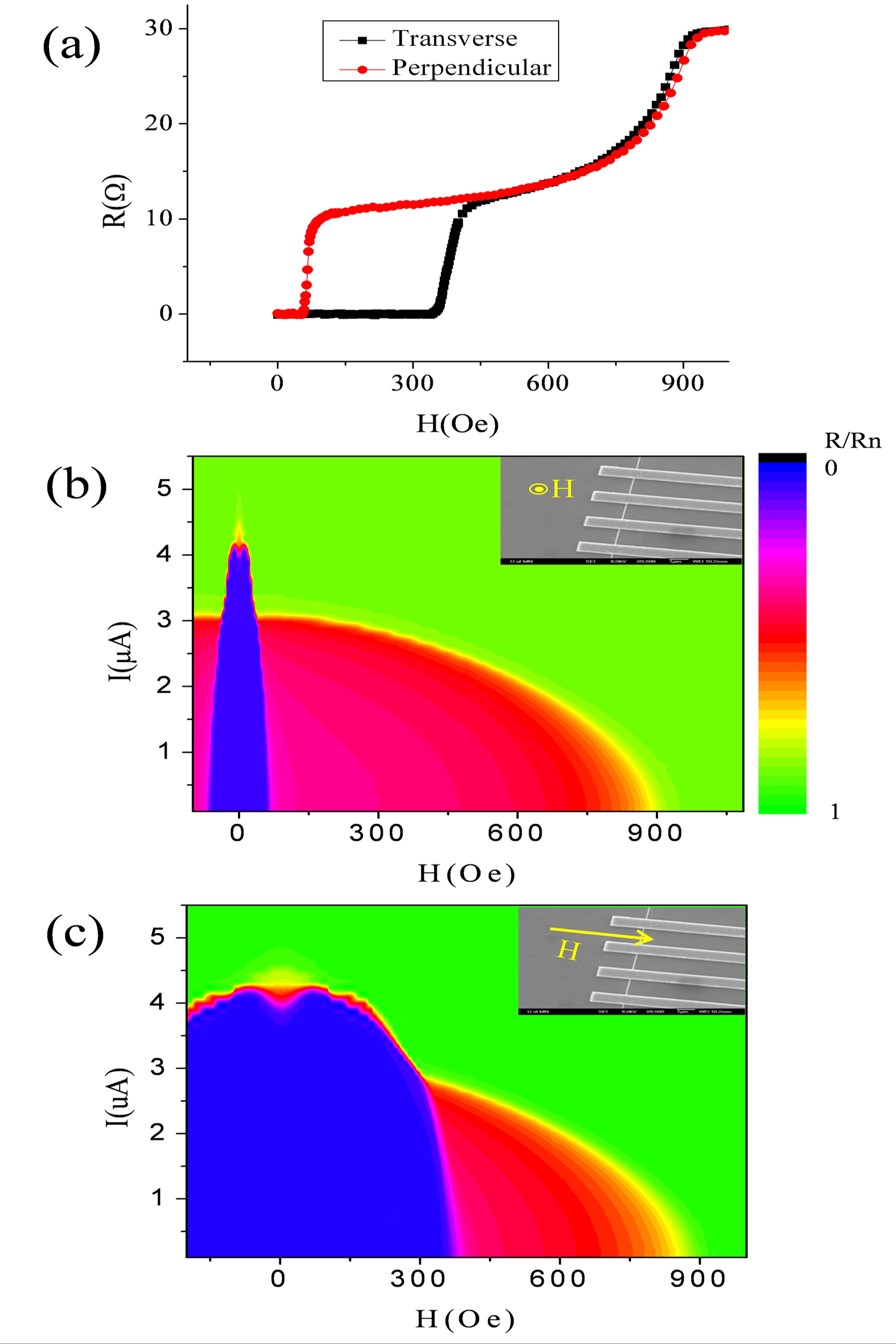}
\par\end{centering}

\caption{\textbf{\label{fig:perpen vs trans}}(Color online) a) Magnetoresistance
of a 1.5 $\mu m$ long wire (Sample B) with different field orientations,
at a temperature 460mK and with a current of 0.4 $\mu A$. b) Phase
diagram for this sample at 460 mK, in a perpendicular field as indicated
in the inset. c) Phase diagram for this sample at 460 mK, in a parallel
field transverse to the axis of the wire as indicated in the inset.}

\end{figure}

\section{\textbf{\textup{\small Field orientation dependence: Boundary effects}}}

One of the most important issues that needs to be addressed is whether
or not these observations represent intrinsic properties of wires,
independent of the boundary electrodes. To this end, measurements
were carried out with different orientations of the applied field
relative to samples, from perpendicular to both the wire and the plane
of the substrate, that was used in the measurements in Fig. \ref{fig:Ic1(T)},
to in the plane of the substrate and transverse to the wire. As labeled
in the inset of Fig. \ref{fig:perpen vs trans}(b), the perpendicular
direction refers to a direction in which the magnetic field is perpendicular
to the plane of the substrate and the transverse direction refers
to field oriented transverse to the wire and in the plane of the substrate.
The false-color phase diagram of a $1.5\mu m$ sample (Sample B) in
this magnetic field direction is shown in Fig. \ref{fig:perpen vs trans}(b).
In this phase diagram, the regime of magnetic field enhanced superconductivity
corresponds to that of high currents ($3.5\mbox{ to }4.5\mu\mbox{A}$),
and low magnetic fields ($-30\mbox{ to }30\mbox{Oe}$). At relative
low currents ($\mbox{I}\lesssim3.0\mu\mbox{A}$), the wire does not
exhibit any sign of an enhancement of superconductivity. The wire
goes through a wide transition region up to 900 Oe, from the zero-resistance
state to the normal state. This represents a two-step transition tuned
by magnetic field. As shown in Fig. \ref{fig:perpen vs trans}(a),
the first step of the transition actually corresponds to the magnetic
field reaching the critical field of the electrodes. %
\begin{figure*}
\begin{centering}
\includegraphics{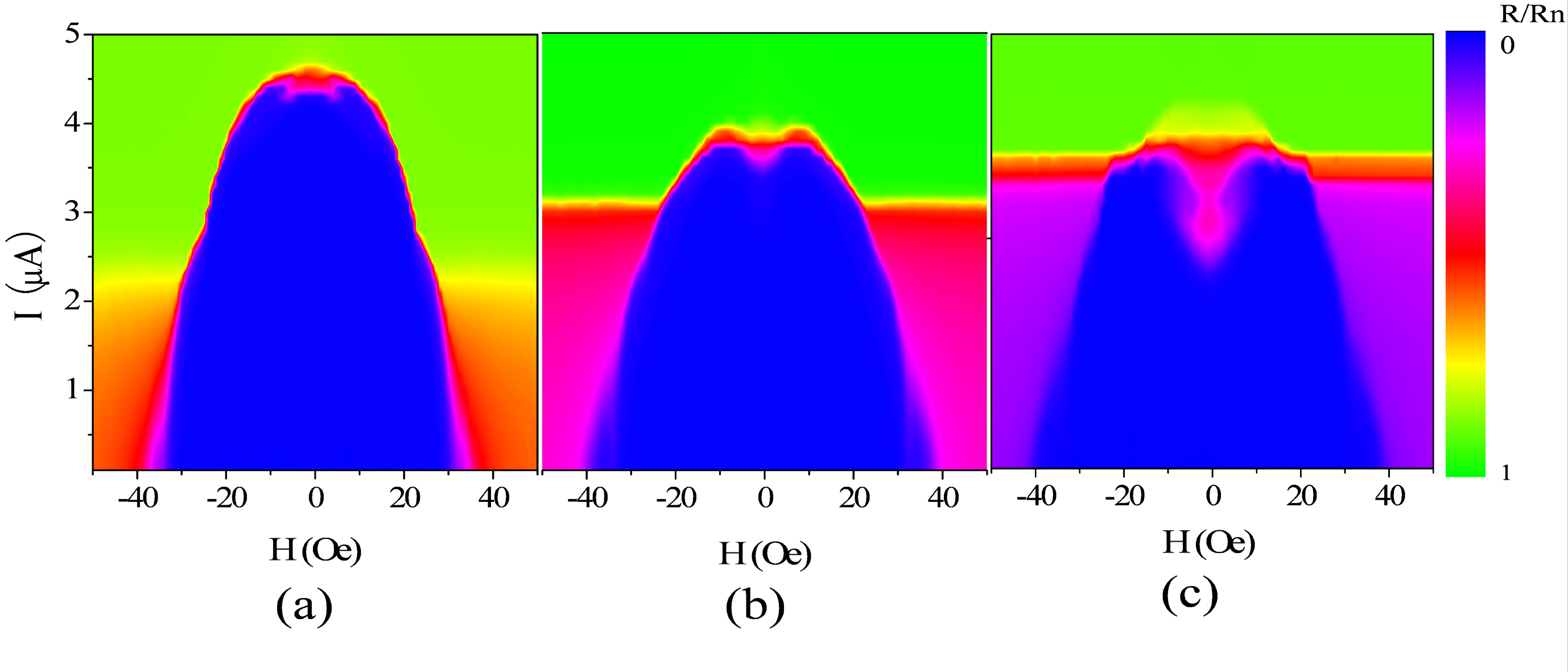}
\par\end{centering}

\caption{\label{fig:RBI length dependence} (Color online)Phase diagrams at
T = 460mK, for wires of different lengths: a) $1\mu m$, b) $2\mu m$,
and c) $4\mu m$. (Samples C, D and E)}

\end{figure*}
The associated resistance change corresponds to a resistance originating
from the proximity effect between the superconducting wire and the
normal electrodes. In addition, it also provides evidence that the
electrodes remain superconducting over the whole regime of re-entrance.
This can be seen by comparing the boundaries of this regime with the
phase boundary associated with the first transition (between the blue
and red regions). Above this field, the system becomes a $NSN$ junction.
The wire resistance increases slowly over a wide a range of magnetic
fields, until it reaches the second step of the transition. At low
temperatures, the magnetic field at this transition is roughly the
critical field of the wire. 

The magnetic field can be switched into the plane of the substrate
and transverse to the wire axis by a $90^{0}$ rotation, as labeled
in the inset of Fig. \ref{fig:perpen vs trans}(c). The height-to-width
ratios are close to unity for the wire but are only around 0.1 for
the electrodes. As a consequence, when the field changes direction
from perpendicular to transverse, it is is always effectively perpendicular
to the wire. For the electrodes, the change of the field orientation
is significant, since the field goes from out-of-plane to in-plane.
To see this, one can compare the previously discussed two-step transition
of the magnetoresistance measurements in both field directions at
a low current, shown in Fig. \ref{fig:perpen vs trans}(a). The critical
field of the electrodes increases by a factor of 5 when the magnetic
field is switched from perpendicular to transverse, while the critical
field of the wire basically remains unchanged. Note that to do this
the sample was brought to room temperature and attached to a different
sample puck in atmosphere. This is responsible for the slight change
in the value of the critical field of the wire at low current between
Fig. \ref{fig:perpen vs trans}(b) and Fig.\ref{fig:perpen vs trans}(c).
Now, focusing on the enhancement regimes in the phase diagrams shown
in Figs. \ref{fig:perpen vs trans}(b) and (c), one can immediately
recognize the expansion of the regime of magnetic field enhanced superconductivity,
when the magnetic field is switched from perpendicular to transverse.
In other words, a higher magnetic field is needed in the transverse
direction to induce reentrance into the superconducting state, compared
with the perpendicular direction. Because the magnetic field remains
perpendicular to the wire, this difference suggests that the observed
enhancement of superconductivity is controlled by processes taking
place in the electrodes in response to the applied magnetic field.
However, what is unchanged is the amplitude of the enhancement, which
can be taken as the increase of the critical current $I_{c1}$. This
should be expected since the response of the wire to magnetic field
is the same in both directions, and the field here is much smaller
than the critical field of the wire itself.

\section{\textbf{\textup{\small Wire length dependence: Two length scales}}}

After establishing that the effect is associated with the boundary
electrodes, we carried out measurements on wires of different lengths
(Sample C: $4\mu m$, D: $2\mu m$ and E: $1\mu m$). In order to
minimize variations associated with fabrication, the wires were produced
in the same process and on the same substrate. The phase diagrams
at 460 mK as a function of current and magnetic field are shown for
each wire in Fig. \ref{fig:RBI length dependence}. By comparing the
enhancement regimes of the three diagrams, the most remarkable feature
is that longer wires exhibit a stronger effect, a larger increase
of the critical current $I_{c1}$. This observation is seemingly counterintuitive
since the enhancement effect has been shown to be a boundary effect.
Naively thinking, the greater the distance to the boundaries, the
less influence they should be expected to exert. One might therefore
expect a longer wire to exhibit a weaker enhancement and eventually
the effect should become negligible for an infinitely long wire. %
\begin{figure}
\begin{centering}
\includegraphics{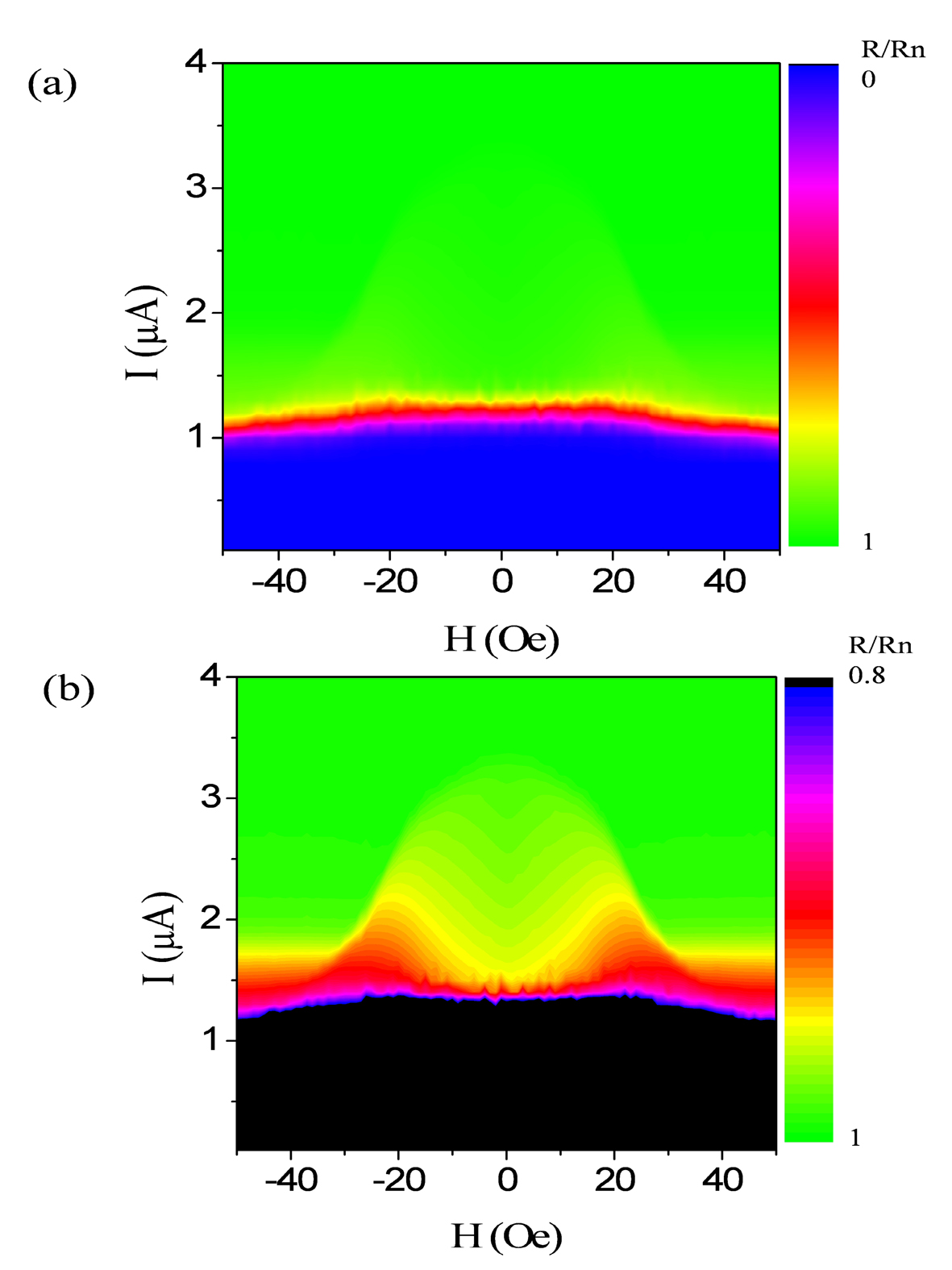}
\par\end{centering}

\caption{\label{fig:RBI 10um}(Color online) Phase diagram of a $10\mu m$
long wire (Sample F) at 460mK. The color scales as $R/R_{n}$: a)
from 0.1to 1, and b) from 0.8 to 1 }

\end{figure}

Further measurements of an even longer wire (Sample F: $10\mu m$),
helped to resolve this issue (this wire was prepared in a separate
process and was thinner than the other three). In its phase diagram
at 460 mK, as shown in Fig. \ref{fig:RBI 10um}(a), the enhancement
effect does seem to disappear. However, as $10\mu m$ is still a finite
length, the segments of the wire near the boundary electrodes should
still be influenced by them, and therefore there should be some remnant
of the enhancement effect in this wire. This can be seen by re-plotting
the phase diagram. Instead of having the colors scale to the full
range of resistances, a new false color plot was generated with the
color scale starting from 80\% of the normal resistance, as shown
in Fig. \ref{fig:RBI 10um}(b). Immediately, the enhancement can be
recognized as the familiar V-shaped structure. However, this structure
can no longer be understood as an increase of the lower critical current
$I_{c1}$. Instead, it is a negative magnetoresistance that is only
a small fraction of the zero field resistance. 

Summarizing the various observations, it is clear that the enhancement
effect becomes weaker in the short wire limit, but also becomes weaker
in the long wire limit. The existence of these two limits strongly
suggests that there are two characteristic length scales that determine
the effect. As we will argue in the following, these should be the
superconducting coherence length and the quasiparticle relaxation
length, two most important length scales for a superconducting system
out of equilibrium.\cite{TinkhamBook}

\section{\textbf{\textup{\small Suppression and stabilization of superconductivity}} }

Before discussing the underlying physical mechanism, there is a fundamental
question that needs to be addressed: Is this effect a true enhancement
of superconductivity by a magnetic field? It has been demonstrated
that applying a small magnetic field can cause an increase in the
values of currents or temperatures at which the wire leaves its zero
resistance state. According to conventional theories of superconductivity,
these currents and temperatures directly relate to the amplitude of
the order parameter.\cite{TinkhamBook} Then, the question becomes
whether or not applying a magnetic field can increase the amplitude
of the order parameter. The answer can be obtained by re-examining
the phase diagrams of wires of different lengths, shown in Fig. \ref{fig:RBI length dependence}.
As mentioned above, these wires were made in the same fabrication
process and on the same substrate. Therefore, they are expected to
have the same amplitude of the order parameter when at the same temperature,
current and magnetic field. Having the superconducting boundaries
included, one would expect the argument above to be valid only for
wires exceeding the superconducting coherence length $\xi$, since
it is the characteristic length scale to which Cooper pairs can coherently
diffuse along the wire. For this reason, we temporarily exclude the
$1\mu m$ wire from the discussion, since its length is on the order
of twice the coherence length. %
\begin{figure}
\begin{centering}
\includegraphics{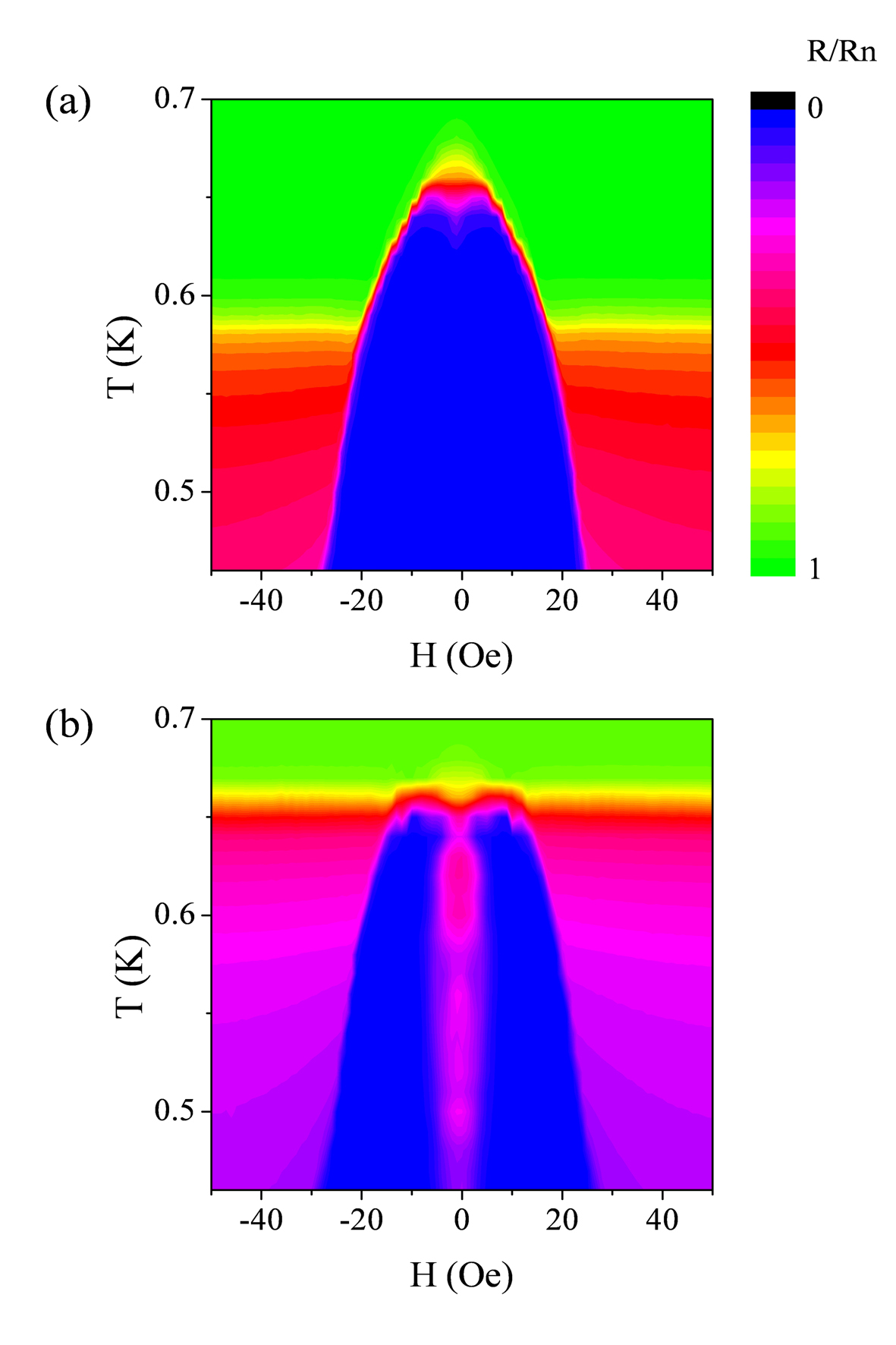}
\par\end{centering}

\caption{\label{fig:RBT length compare}(Color online) Phase diagrams at a
current of $2.5\mu A$ of wires of different lengths: a) $L=2\mu m$,
and b) $L=4\mu m$. (Samples D and E)}

\end{figure}

A comparison of the phase diagrams of the $4\mu m$ and the $2\mu m$
sample can be made. We first consider the values of $I_{c1}$ at which
the wires leave their superconducting states in zero field. It is
$\sim3.5\mu A$ for the $2\mu m$ wire and $\sim2.5\mu A$ for the
$4\mu m$ wire, a difference of approximately $40\%$. On the other
hand, if one compares the maximum value of $I_{c1}$ in a small magnetic
field, one can see that the two wires share almost the same value
$\sim3.7\mu A$. This is more evident from a similar comparison of
the critical temperatures $T_{c1}$ at which wires leave the zero
resistance state at a certain current. The phase diagrams of these
two wires at $I=2.5\mu A$, as a function of temperature and magnetic
field, are shown in Fig. \ref{fig:RBT length compare}. Once again,
one can see that the maximum values of $T_{c1}$ are almost the same
$\sim0.65K$ for both wires, but in zero field $T_{c1}$ differs by
$\sim40\%$ ($T_{c1}\sim0.64K$ for the $2\mu m$ wire and $T_{c1}\sim0.46K$
for the $4\mu m$ wire). 

Combining the arguments regarding similar superconducting properties
between the two samples, it is therefore more appropriate to treat
the effect in two steps. First, superconductivity, or more accurately
the zero resistance state of the wire is suppressed in zero magnetic
field by the applied current. Second, a small magnetic field can induce
a \textit{recovery }or\textit{ stabilization} of the suppressed superconductivity. 

\begin{figure}
\begin{centering}
\includegraphics{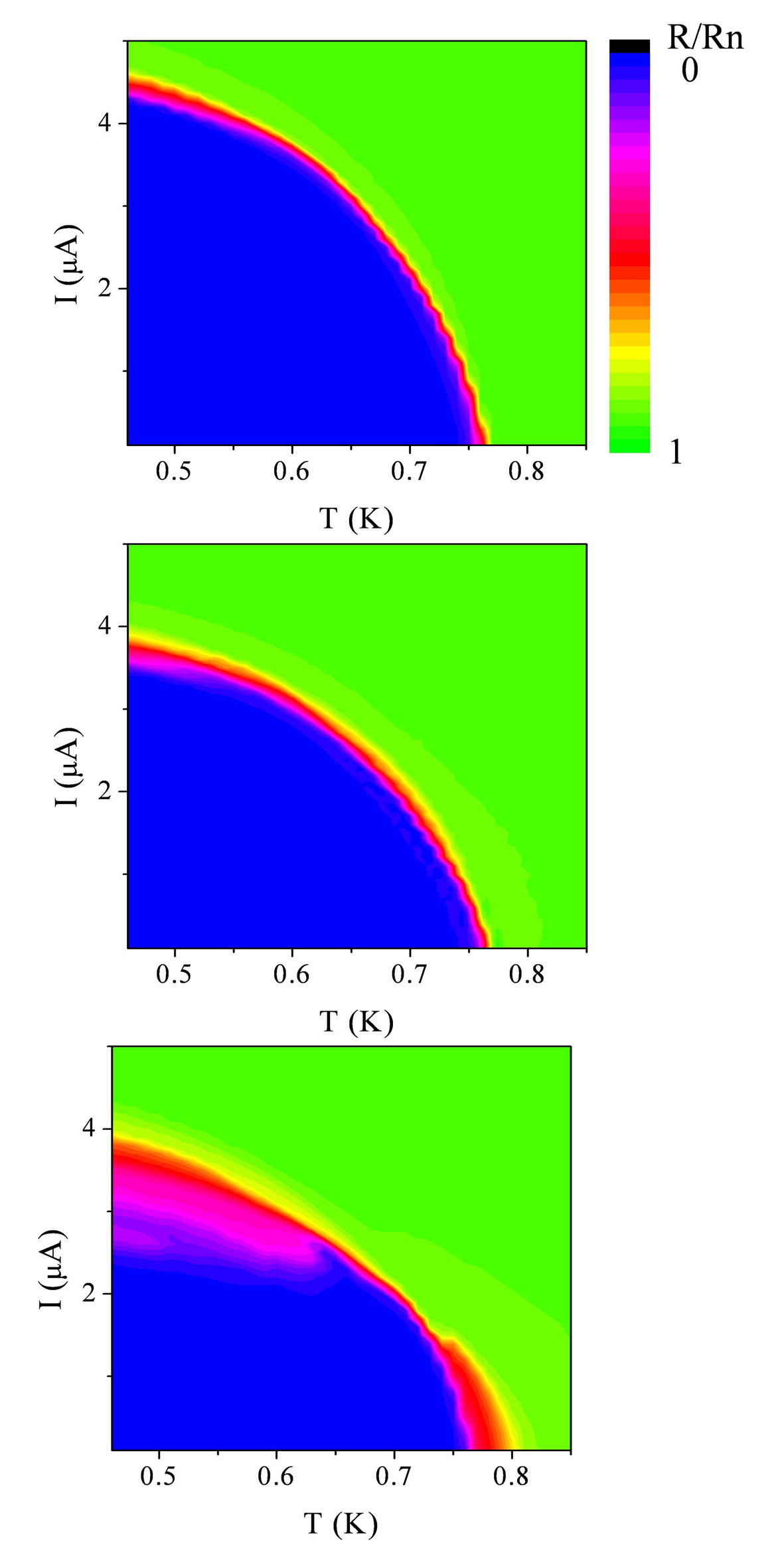}
\par\end{centering}

\caption{\textbf{\label{fig:RT 0Oe length}}(Color online) Phase diagrams at
zero field for wires of different lengths a) $1\mu m$, b) $2\mu m$,
and c) $4\mu m$\textbf{. }(Samples C, D and E) }

\end{figure}

It is therefore useful to examine the phase diagrams for these three
wires in zero field, shown in Fig. \ref{fig:RT 0Oe length}.%
 The suppression is exhibited as broadened transition regions. Cutting
slices out of zero-field phase diagrams at $T=0.46K$ for example,
one can now understand the phase diagrams of Fig. \ref{fig:RBI length dependence},
from the point of view of a recovery of the suppressed zero resistance
state when applying a magnetic field. A longer wire would be expected
to have a stronger {}``enhancement'' effect since it would exhibit
a stronger suppression by current in zero field. A similar argument
applies to differences between the diagrams in Fig. \ref{fig:RBT length compare},
for the increase of $T_{c1}$ at $2.5\mu A$ in fields.

\section{\textbf{\textup{\small Discussion}}}

Even though it is evident that the observed effect is a recovery of
the suppressed superconductivity, its actual physical mechanism remains
unclear. In the following, we will present some phenomenological explanations.
Following the two-step interpretation, we first consider the suppression
of the superconductivity of the wire. At zero field, the suppression
of the superconductivity is exhibited as a resistive state over a
wide range of applied currents. It is therefore essential to understand
the origin of the resistance in the transition regime, as well as
the shoulder-like structure. A superconducting wire with non-zero
order parameter can acquire a non-zero resistance through several
mechanisms. The first one is associated with the penetration of the
electric field, when the wire is connected to normal electrodes. This
mechanism apparently does not apply here since the electrodes remain
superconducting over the whole range of reentrance, as discussed in
Section III.

The second mechanism is associated with the formation of phase slip
centers. In this case, instead of abruptly transitioning from the
superconducting to the normal state at the depairing current, the
wire remains resistive over a range of currents. A theory of the current-driven
destruction of superconductivity of wires in the quasi-one-dimensional
limit was was developed by Kramer and Baratoff.\cite{Kramer1977}
Their numerical calculations, based on the time-dependent GL equations,
demonstrated that the current-driven destruction of superconductivity
in one dimension is associated with two currents $I_{min}$ and $I_{max}$.
For $I<I_{min}$, the superconducting state is stable. For $I>I_{max}$,
the normal state is stable. For $I_{min}<I<I_{max}$, the system remains
superconducting while becoming resistive due to phase slip processes.
This region corresponds to the transition regime. Since $I_{max}$
can be roughly taken as the depairing current, this calculation suggests
that quasi-one-dimensional superconductors can lose their zero resistance
state at currents lower than the depairing current. This resembles
the suppression of superconductivity discussed here. Accordingly,
$I_{min}$ and $I_{max}$ correspond to $I_{c1}$ and $I_{c2}$ in
the present study. 

For wires of lengths less than the quasiparticle relaxation length
and connected to superconducting electrodes, the location of the phase
slip center would most likely be at the midpoint of the wire. The
observation that shorter wires have a weaker suppression can therefore
be explained as coming from their superconductivity being more strongly
supported, or the phase slip processes being more strongly suppressed,
by the superconducting boundaries. This has been treated in a theoretical
study of the conditions for the occurrence of phase slip centers.\cite{Michotte2004}
The critical current $j_{c1}$ at which phase slips start to occur
is obtained by comparing the relaxation rates of the amplitude and
phase of the order parameter. With superconducting boundaries, 

\begin{equation}
j_{c1}\thicksim\frac{j_{o}}{\tanh(L/2\Lambda_{Q})}\label{eq:2}\end{equation}
Here $\Lambda_{Q}$ is the quasiparticle relaxation length and $j{}_{0}=c\Phi_{0}/8\pi^{2}\Lambda^{2}\xi$
(the GL critical current $\sim0.385j_{0}$). One therefore can see
that $j_{c1}$ decreases until saturating when the wire length exceeds
a certain value. This corresponds to the case in which the phase slip
center is out of the range of the boundary, with respect to the exchange
of quasiparticles and Cooper pairs. For wires of lengths approaching
the coherence length, it is more appropriate to treat the system as
a S-c-S junction (S stands for superconductor, c stands for constriction).
In this case, the system will have a higher critical current since
it now can withstand a higher phase gradient $\sim1/L$ instead of
$\sim1/\xi$ for longer wires. This has been observed as a high critical
current for the $1\mu m$ wire in our study, as shown in Fig. \ref{fig:RBI length dependence}(a).\cite{TinkhamBook} 

The coupling between the wire and superconducting electrodes can also
help us to understand the shoulder-like structure in $R-I$ curves.
This structure suggests a crossover between two different mechanisms
associated with resistance in the transition region. In the lower
part, the wire is superconducting and the voltage originates from
the phase fluctuations at phase slip centers as discussed previously.
In the upper part, the wire has been driven normal, and the reduction
relative to the normal resistance comes from the proximity effect
with the superconducting electrodes. Accordingly, the lower critical
current, $I_{c1}$, is the current at which the order parameter at
the midpoint of the wire has become weak enough to be destroyed by
the fluctuations. 

The second issue is how the application of a magnetic field can lead
to the recovery of the suppressed superconductivity. The existing
theoretical models are associated with the polarization of spin fluctuations,
negative Josephson coupling, reduction of quasiparticle relaxation
lengths and the dissipation dampening of phase slips.\cite{Rogachev2006,Spivak1991,Arutyunov2008,Vodolazov2007,Zaikin1998}
As discussed in Ref.16, the first two models cannot be applied to
our results. In the following, our discussion will focus on the last
two. 

$\Lambda_{Q}$ is known to be reduced upon the application of a magnetic
field.\cite{Clarke} As phase slips are accompanied by quasiparticle
relaxation processes, theoretical studies have shown that this reduction
can effectively lead to {}``enhancements'' of superconductivity
in one dimension, either as a negative magnetoresistance\cite{Arutyunov2008}
or as an increase of the critical current.\cite{Vodolazov2007} In
particular, the critical current $j_{c1}$ has been predicted to increase
upon the application of a small magnetic field, and it is especially
pronounced for weak superconductors such as Al and Zn. Larger fields,
on the other hand, will lead to a decrease of $j_{c1}$. These increases
and decreases originate from the fact that magnetic fields not only
reduce $\Lambda_{Q}$ but the order parameter as well.\cite{Vodolazov2007}
This prediction is in good agreement with the observations in the
present work, however, several problems still exist with the application
of the model to the experimental results. The discussion of the behavior
of the relaxation length only considers its variation with field due
to changes of the order parameter of the wire. In contrast, the dependence
of the observed effect on field orientation demonstrates that the
superconducting boundary electrodes play a major role. In addition,
the model is based on the time-dependent GL equation, which is not
valid in the low temperature regime. 

It has also been suggested recently that the small magnetic fields
used here cannot appreciably change the value of $\Lambda_{Q}$. However,
this does not rule out the possibility of changes in the order parameter
in the leads being the cause of the effect. When the order parameter
in the leads is suppressed the diffusion of quasiparticles into the
wire can become stronger. This could help explain the results seen
here and should be pursued further.\cite{Vodolazov2010} 

As discussed above, the wire resistance in the transition regime below
the shoulder is associated with phase slips driven by fluctuations.
However, the nature of these fluctuations is unclear for the nanowires
studied in the present work. Numerical fits of various models of the
temperature dependence of the wire resistance have been carried out.
In the low current limit, reasonable fits of thermal activated phase
slip models can be obtained.\cite{Langer1967,McCumber1970} 

\begin{equation}
R=R_{Q}\frac{\hbar\Omega(T)}{k_{B}T}e^{-\Delta F(T)/k_{B}T}\label{eq:TAPS}\end{equation}
Here, $R_{Q}=h/4e^{2}$ is the quantum resistance for Cooper pairs,
$\Omega(T)$ is the attempt frequency, and $\Delta F$ is the energy
cost of nucleating a phase slip:

\begin{equation}
\Omega(T)=\frac{L}{\xi(T)}\left(\frac{\Delta F(T)}{k_{B}T}\right)^{1/2}\frac{8k_{B}(T_{c}-T)}{\pi\hbar}\label{eq:attemptF}\end{equation}

\begin{equation}
\Delta F(T)=\frac{8\sqrt{2}}{3}\frac{H_{C}^{2}(T)}{8\pi}A\xi(T)\label{eq:energyBarrier}\end{equation}
Here $H_{c}(T)$ is the thermodynamic critical field and $\xi(T)$
is the GL coherence length. Increasing current induces a broadening
of the transition such that all the existing models, including those
which treat quantum phase slips,\cite{Giordano1988,Zaikin1997} fail
to fit the data. This does not exclude the possibility of quantum
phase slips. 

\begin{figure}
\begin{centering}
\includegraphics{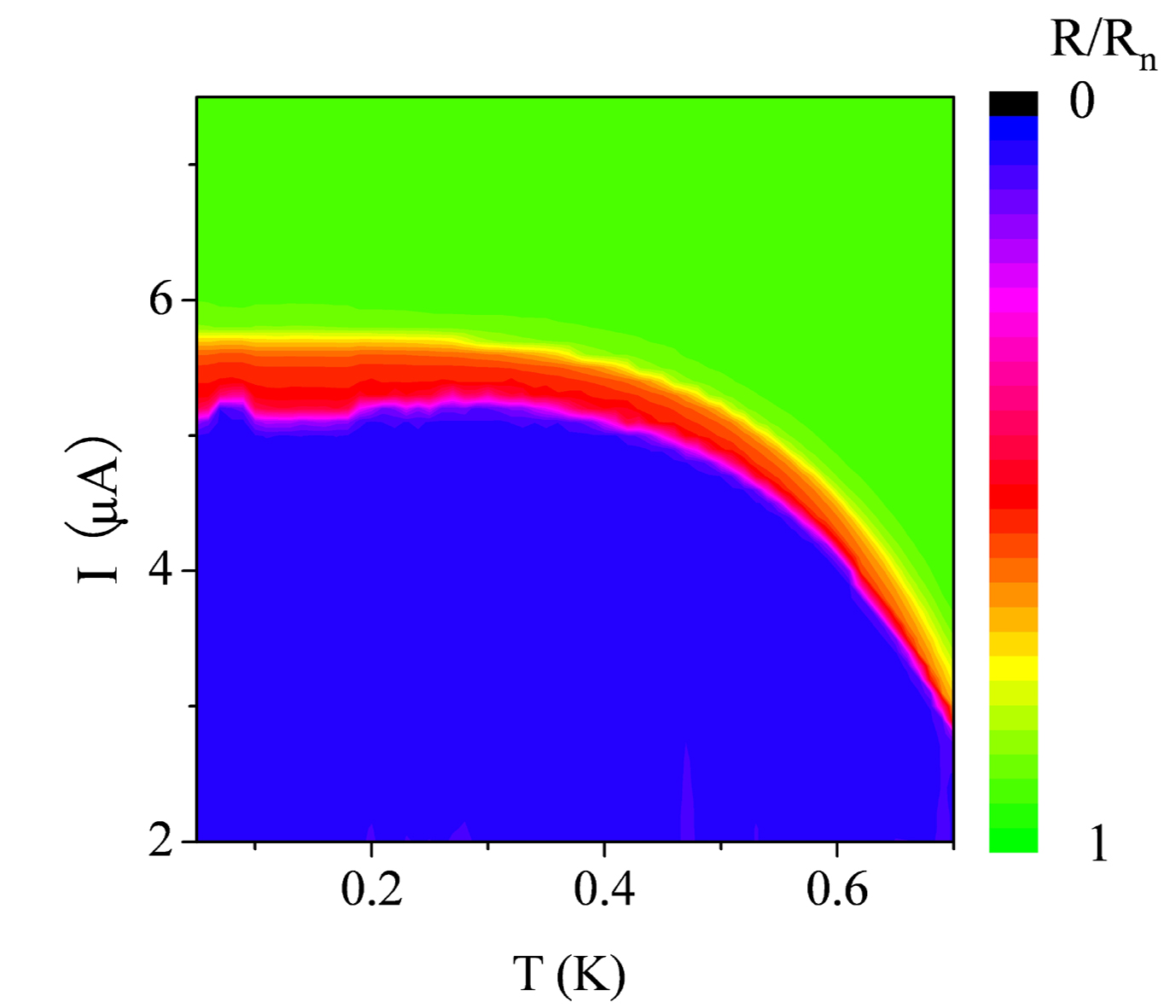}
\par\end{centering}

\caption{\label{fig:RT 0Oe lowT}(Color online) Phase diagram of a $1.5\mu m$
wire (Sample G) in zero field, with the temperature range extended
down to 50mK. The while line labels the position of the shoulder in
$R-I$ curves. }

\end{figure}

In Fig.\ref{fig:RT 0Oe lowT}, we demonstrate a zero-field phase diagram
of a $1.5\mu m$ wire (Sample G), with the position of the shoulder
in the $R-I$ data labeled as the white line.%
 It is clear that the transition region below the shoulder becomes
more pronounced at lower temperature, and persists towards zero temperature.
As thermal fluctuations are negligible in this temperature range,
the result suggests that phase slips in this case might be driven
by quantum fluctuations, even though the data cannot fit by existing
quantum phase slip (QPS) models.

If the resistance seen in the transition region is driven by QPSs,
the field-enhanced superconductivity could then be explained through
the interplay between quantum fluctuations and dissipation associated
with the increased population of quasiparticles generated upon the
application of a magnetic field. Theories of the interplay between
dissipation and phase slips driven by quantum fluctuations in superconducting
nanowires have been presented.\cite{Zaikin1997,Refael2007,Zaikin1998}
These exploit analogies between nanowires and Josephson junctions.\cite{Schmid1983,Chakravarty1982}
Dissipation can dampen QPSs and therefore stabilize superconducting
long-range order. For a wire connected to electrodes, theoretical
studies have shown that dissipation from the external environment
can also dampen phase slips.\cite{Buchler2004,Fu2006} 

When a magnetic field is applied, it will affect both the phase slips
and the dissipation. This was discussed in a calculation of QPS-induced
wire resistances: 

\begin{equation}
R(T)\propto\widetilde{y}(T\widetilde{\tau_{0}})^{2\gamma-2}\label{eq:QPS}\end{equation}
Here $\gamma\sim\sigma_{QP},$ where $\sigma_{QP}$ is the quasiparticle
conductance. The effective fugacity of the interacting QPSs, $\tilde{y}$,
is proportional to the rate of QPSs.\cite{Zaikin1998} An applied
magnetic field suppresses the order parameter and increases the rate
of QPSs. This leads to an increase of the wire resistance. On the
other hand, it will enhance the dissipative term $\gamma$ with an
increased density of quasiparticles. This increase can be exponentially
large at low temperatures compared with the power-law change of the
QPS rate. Therefore, a negative magnetoresistance is expected in the
low field regime. 

Applying these ideas to the present experiments, a phenomenological
scenario can be constructed. Increasing the current results in manifestation
of the phase slips, which broaden the resistive transition. When a
magnetic field is applied, the suppression of the order parameter
of the electrodes results in the generation of a large number of quasiparticles.
These quasiparticles, when traversing the wire, dampen the phase slip
processes that produce resistance, resulting in the wire recovering
its zero resistance state. 

It is also important to address the issue of heating. At high currents,
heating has been known to strongly modify the transition of superconducting
wires. The consequence of heating is that a hot spot will be formed
and quickly expand along the wire.\cite{Skocpol1974,Tinkham2003}
In this case, the wire will directly jump to its normal state and
remain normal unless the current is lowered. In this case, the I-V
curves are expected to be hysteretic. However, in the present study,
both the current and magnetic field driven transitions are reversible
without any hysteresis. It indicates the samples are not only in the
overdamped regime, but also are sufficiently cooled by their electrical
connections and the substrate. This is possibly associated with the
fact that the cross-sectional areas are relatively large and resistances
are small in the samples. In addition, we exclude the possibility
that the magnetic-field-stabilized superconductivity comes from the
enhanced thermal conductivity of both the wire and the electrodes
in the magnetic field. This exclusion has been discussed in detail
in Ref.16, based on the fact that different parts of the transition
regime respond to magnetic field differently.

\section{\textbf{\textup{\small Conclusion}}}

When the wire length is short, with a length on the order of the superconducting
coherence length, the\textit{ suppression} of superconductivity by
the applied current is weak. In this case, the superconductivity of
the wire is strongly supported by the superconducting electrodes.
The Cooper pairs can easily propagate coherently over the whole length
of the wire and phase slips are rare. In the other limit, when the
wire length approaches the quasiparticle relaxation length, the \textit{stabilization
}of superconductivity by magnetic field is weak. In this case, quasiparticles
from the electrodes need to diffuse a long distance in order to reach
the phase slip center - the midpoint of the wire. As a consequence
there is a high probability for them to be converted into Cooper pairs.
The effect of quasiparticles in dampening phase slip processes is
therefore limited. This is why the observed effect is weak when the
wire is too short or too long. The recovery of superconductivity is
strongest for wires of intermediate lengths, shorter than the quasiparticle
relaxation length, but longer than the coherence length. 

Although the above is a phenomenological explanation of the effect,
it is developed from a set of inferences based on experimental observations
and theories not specific to the detailed experimental configuration.
A formal theory that can be compared in detail with the experimental
results would be needed to fully understand the observed phenomena.

\begin{center}
\textbf{ACKNOWLEDGMENT}
\par\end{center}

The authors thank Alex Kamenev for useful discussions. This work was
supported by the U.S. Department of Energy under Grant No. DE-FG02-
02ER46004 and by the National Science Foundation under Grant No. NSF/DMR-0854752.
Part of this work was carried out at the University of Minnesota Characterization
Facility and the Nanofabrication Center, which receive partial support
from the NSF through the NNIN program.

\end{document}